# Local Electromagnetic Fields Enable Fast Redox Sensing by Physically Accelerating Cysteine Oxidation

James N. Cobley

The University of Dundee, Dundee, Scotland, UK.

**Correspondence**: j_cobley@yahoo.com or jcobley001@dundee.ac.uk

**ORCID**: https://orcid.org/0000-0001-5505-7424

**Abstract**

Hydrogen peroxide ($H_2O_2$) oxidises cysteine residues to control protein function, yet bulk rate constants predict hours for changes that occur in cells in seconds. Here, this work shows that local electromagnetic fields (EMFs)—ubiquitous in proteins, membranes and nanodomains—can lawfully modulate the Eyring barrier and orientate reactants, accelerating cysteine oxidation without changing the underlying chemistry. Embedding a field term into the Eyring expression, $k(E) = \kappa_0 \frac{k_B T}{h} exp[-(\Delta G_0^{\ddagger} - \Delta\mu\, E - \frac{1}{2}\Delta\alpha E^2)/RT]$, demonstrated that plausible local EMFs ($10^8 - 10^9\, V \cdot m^{-1}$) with realistic dipole changes ($\Delta\mu \approx 10 - 20\, D$), accelerate rate constants by orders of magnitude. This local acceleration reconciles the discrepancy between predicted vs. observed rates of $H_2O_2$-mediated cysteine oxidation. The framework generates falsifiable predictions—vibrational Stark readouts in thiolate–peroxide complexes should fall within 1–10 cm⁻¹ per (M V cm⁻¹)—and reframes rate-constants as mutable, field-condition parameters. Cysteine redox sensing is fast not because the chemistry is exotic, but because the physics is local.

**Main**

Cysteine oxidation by hydrogen peroxide ($H_2O_2$) underlies redox control across metabolism, signalling, and stress adaptation[1]. However, empirical observations of cysteine redox sensing—often within seconds and minutes—are not predicted by bulk kinetics[2]. Representative kinetic values for most cysteines[3] ($k_2 \approx 10\, M^{-1}s^{-1}$), predict hours to achieve 5%-oxidation of a single residue in a cell at plausible sub-micromolar [$H_2O_2$] levels (**Figure 1A**), even assuming no cysteine reduction and constant $H_2O_2$ replenishment.

To reach 5%-oxidation within 5-min, either 17 $\mu$M $H_2O_2$ would need to be supplied each second or the reaction would need to be accelerated $\sim 17-$fold to $k_2 \approx 171\, M^{-1}s^{-1}$ (**Figure 1B-C**). In a cell with a volume of 3 picolitres, 17 $\mu$M maps to $\approx 3.1 \times 10^7$ [$H_2O_2$] molecules, compared with just $\approx 1.8 \times 10^3$ at plausible "baseline" of 1 nM. Sustaining a $\sim 1.7 \times 10^4$-fold [$H_2O_2$] expansion during redox sensing seems improbable. Fast cysteine redox sensing still persists even when indirect kinetic accelerators—like the transfer of electrons from $H_2O_2$ to target cysteines via fast-reacting enzymes[4] —are disabled[5].

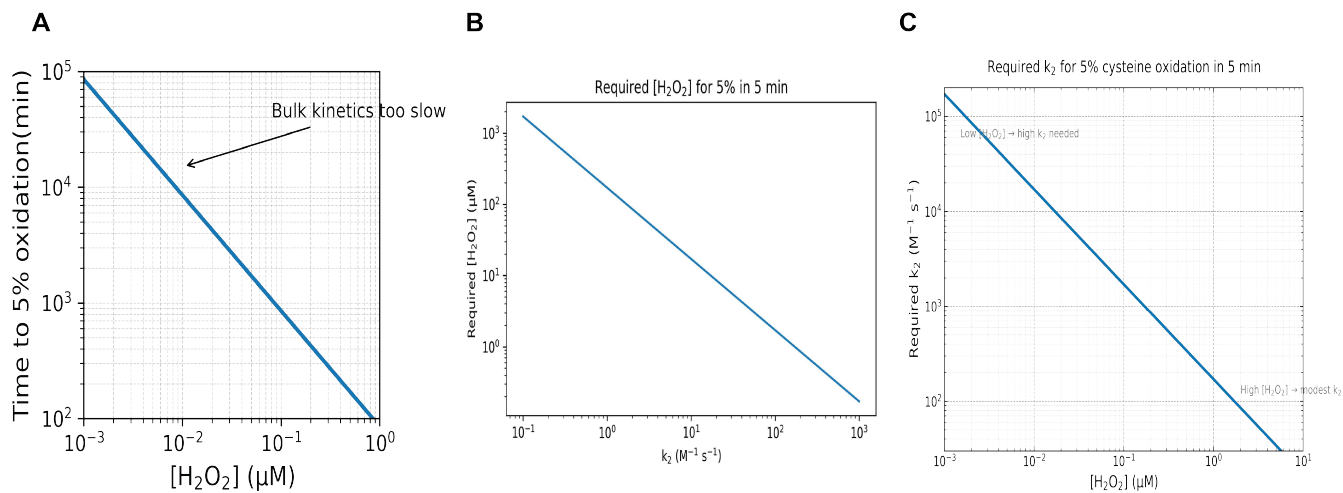

**Figure 1. Bulk H₂O₂ mediated cysteine oxidation is too slow for fast redox sensing. A.** The time to reach 5% cysteine oxidation of one site in a cell at different [H₂O₂] concentrations. **B.** The required [H2O2] needed to oxidised 5% of a cysteine residue at different kinetic rates. **C.** The required kinetic rate for 5% cysteine oxidation in 5-min.

The pronounced mismatch between predicted vs. observed redox sensing timeframes suggests something fundamental is missing from the equations themselves: local electromagnetic fields (EMFs). Since the EMF physically mediates the action of the electric and magnetic forces on charged or polar molecules, it appears explicitly in the Arrhenius and Eyring[6] formulations

$$k = \kappa \frac{k_B T}{h} exp(-\Delta G^{\ddagger}/RT)$$

where both the activation barrier $\Delta G\ddagger$ and the prefactor $\kappa$ are field-sensitive. In the test-tube environments from which typical rate constants ($k_2 \approx 10\ M^{-1}s^{-1}$) were derived[7], local EMFs are effectively zero ($E = 0$) due to screening by the bulk isotropic solvent. However, the $E = 0$ assumption fails inside cells, where proteins, membranes, and nanodomains generate intense local EMFs (e.g., $E \sim 10^7\ V\ m^{-1}$ across the inner mitochondrial membrane[8]).

Consistent with this, local EMF were modelled near a negatively charged membrane ($\varphi_0 \approx -70\ mV$) under physiological ionic strength ($150\ mM$) and a low-dielectric ($\varepsilon \approx 10$) protein microenvironment (**Extended Methods**). Within 1–5 nm of the surface, the field decayed exponentially from ~$10^8$ to $10^7$ V m⁻¹. When the cysteine was surrounded by positively charged residues, the combined effect of membrane and residue charges yielded total normal fields of 1–3 × 10⁸ V m⁻¹ (**Figure 2A**), within a biologically plausible range[9–11].

These local EMFs are described by the electro-quasistatic limit of Maxwell's equations, appropriate for nanometre–microsecond regimes where magnetic induction and displacement currents are negligible. In aqueous and membrane media ($\sigma \approx 1\ S\ m^{-1}, \varepsilon \approx 80, \varepsilon_0$), the conduction–displacement crossover occurs near $10^8$ Hz—five to seven orders of magnitude above the kilohertz-scale frequencies of ionic or membrane potential fluctuations in cells. Consequently, local EMFs in biological structures, even in dynamic environments, can be regarded as quasi-static and irrotational ($\nabla \times E \approx 0$), governed by the scalar potential $\varphi$ through $\nabla \cdot (\varepsilon \nabla \varphi) = -\rho_f$.

Despite their potential to lawfully accelerate kinetic rates, the formalism for integrating local EMFs into the equations of redox sensing is missing. To establish this missing formalism, the local EMF was explicitly embedded into the Eyring equation

$$k(E) = \kappa_0 \frac{k_B T}{h} \exp\left[-\frac{(\Delta G_0^\ddagger - \Delta\mu\, E - \frac{1}{2}\Delta\alpha E^2)}{RT}\right]$$

where $\Delta\mu$ and $\Delta\alpha$ are the dipole and polarizability changes between reactants and the transition state (**Extended Methods**). The linear Stark term ($\Delta\mu E$) and the quadratic term ($\frac{1}{2}\Delta\alpha E^2$) shift the barrier height, while $\kappa_0$ gains a multiplicative orientation factor $f_{orient} \approx \langle \cos\theta \rangle \exp(\Delta\mu E/k_B T)$. The combined effect mathematically yields an exponential rate amplification proportional to $exp[(\Delta\mu E + \frac{1}{2}\Delta\alpha E^2)/RT]$.

Computationally instantiating this field-integrated formalism over a range of plausible local EMFs[9–11] ($10^7 - 3 \times 10^8\ V\ m^{-1}$) and dielectric focusing factors ($F\varepsilon = 1 - 5$) revealed an exponential acceleration of the apparent kinetic rate constant by $10 - 10^3$-fold (**Figure 2B-C**). Even a moderate 30-fold enhancement of the rate constant enables 5%-cysteine oxidation to be achieved in 28-min and 2.8-min at 0.1 and $1\mu M$ $H_2O_2$, respectively (**Figure 2D**). A 100-fold enhancement achieves 5%-oxidation in 8.5-min at 0.1 $\mu M$ $H_2O_2$, a plausible signalling concentration and timeframe[12]. Hence, local EMFs reconcile slow bulk $H_2O_2$ kinetics with fast cysteine redox sensing.

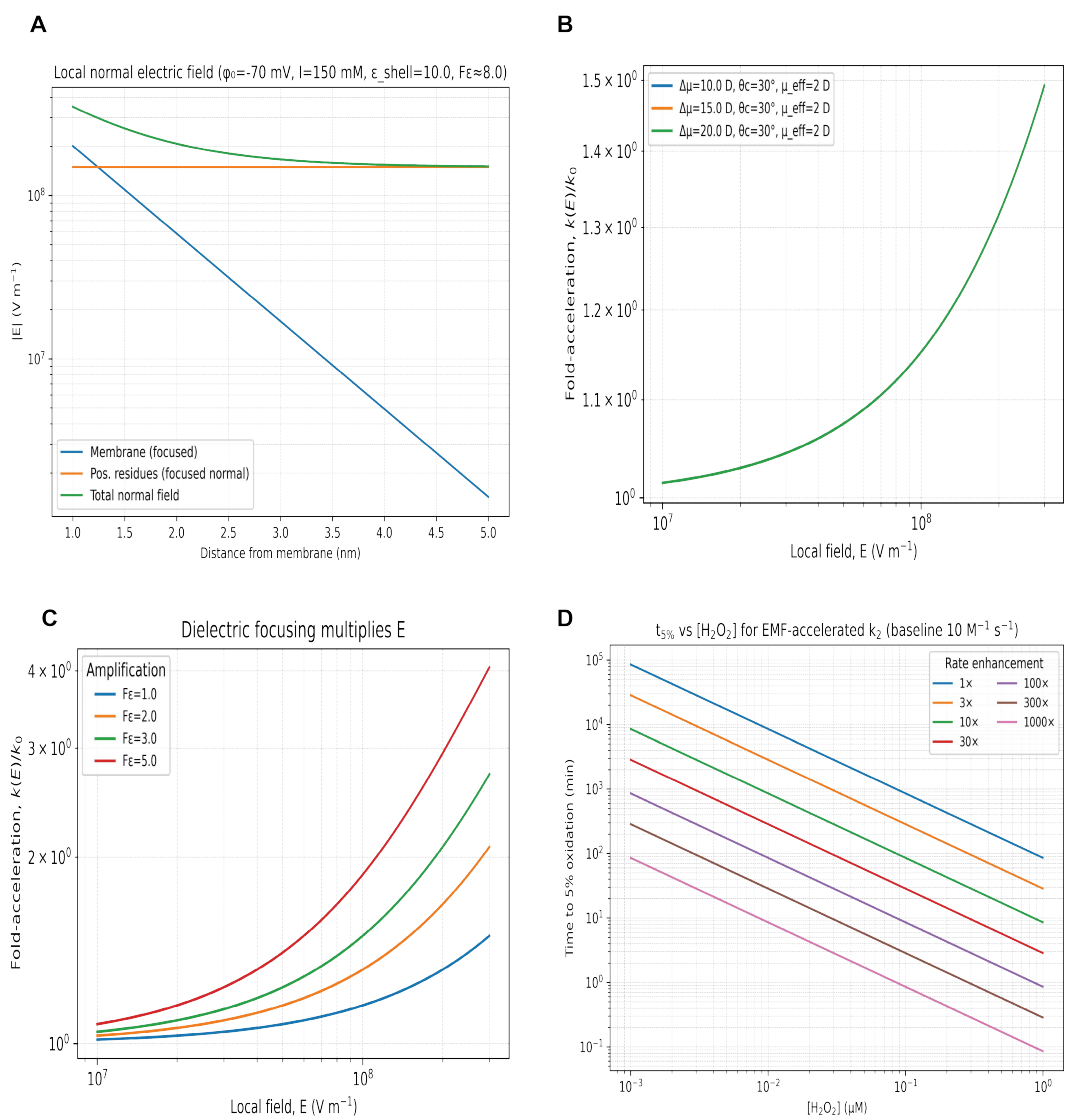

**Figure 2. Local EMFs accelerate cysteine redox sensing. A.** Local electric field magnitudes near a −70 mV membrane potential at physiological ionic strength (150 mM) and in a low-dielectric (ε ≈ 10) protein microenvironment. Within 1–5 nm of the surface, total normal fields reach $10^7$–$10^8$ V m$^{-1}$ when contributions from positively charged residues are included. **B.** Field-dependent exponential acceleration predicted by the field-integrated Eyring formalism, $k(E)/k_0 = \exp[(\Delta\mu E + \frac{1}{2}\Delta\alpha E^2)/RT]$, for dipole changes ($\Delta\mu$ = 10–20 D). Plausible local EMFs ($10^7$–$3 \times 10^8$ V m$^{-1}$) increase apparent rate constants by 10–$10^3$-fold. **C.** Dielectric focusing further amplifies the local field by factors of 1–5 as ε decreases, producing second-order rate enhancements proportional to $E^2$. **D.** Predicted impact of field-driven rate enhancement on the time to 5 % oxidation across $H_2O_2$ concentrations. Even modest field strengths (≈$10^8$ V m$^{-1}$) reconcile slow bulk kinetics with the rapid oxidation times observed in cells.

When an $H_2O_2$ molecule enters the field domain of a cysteine, the local EMF immediately distorts its O–O bond, shifting electron density toward one oxygen atom and inducing a dipole along the bond axis. This field-induced pre-polarisation renders one oxygen atom more electrophilic, preconditioning it for nucleophilic attack by the thiolate. Once the attack begins, the local EMF couples to the evolving charge distribution along the reaction coordinate. The transition from thiolate ($S^-$) to sulfenic acid ($SOH$) involves electron transfer and bond polarization, generating a transient dipole change ($\Delta\mu$) and polarizability shift ($\Delta\alpha$). The field performs work on this dipole according to

$$\Delta G^{\ddagger}(E) = \Delta G_0^{\ddagger} - \Delta\mu_{\parallel} E - \frac{1}{2}\Delta\alpha E^2,$$

reducing the activation barrier by several $k_B T$ for plausible $E \approx 10^8 - 10^9\ V\ m^{-1}$ and $\Delta\mu \approx 10 - 20\ D$. Simultaneously, the field orients the reactants within a reaction cone, increasing the fraction of productive collisions. The field also stabilizes the otherwise unfavourable hydroxide leaving group. In the transition state, partial O–O cleavage and S–O bond formation create charge separation ($\delta^-$ on the departing –OH and $\delta^+$ along the peroxide axis). A field aligned with this coordinate performs electrostatic work ($\Delta G_{field} \approx qE\Delta r$), effectively replacing solvent stabilization by directly lowering the energy of the emerging anion. For $E \approx 10^8 - 10^9\ V\ m^{-1}$ and a 1 Å charge displacement, $\Delta G_{field}$ corresponds to $\approx 1 - 10\ kJ\ mol^{-1}$ (0.4 − 5 )≈1–10 kJ mol$^{-1}$ (0.5–4 $k_B T$), sufficient to enhance the rate constant by one to two orders of magnitude. Hence, the local EMF functions as an intrinsic electrostatic catalyst—polarising the O–O bond, stabilising the leaving –OH groups, and aligning electron flow along the reaction coordinate.

The framework yields falsifiable tests. Vibrational Stark shifts[13] in thiolate–peroxide complexes should fall within 1–10 cm$^{-1}$ per M V cm$^{-1}$, confirming field strengths of $10^7$–$10^8$ V m$^{-1}$. Encapsulating cysteines in lower-permittivity nanodomains (condensates) should accelerate oxidation rates $\propto E^2$ via dielectric tuning. Substituting charged residues near reactive cysteines should modulate redox kinetics predictably. Finally, inert polymers that compress solvent and reduce dielectric shielding accelerate oxidation. The latter has already been experimentally observed[14]. These tests transform a conceptual insight into an experimental program.

Rate acceleration can only occur when the *reactive peroxide molecule physically occupies the region where the local field is non-zero at the cysteine reactive coordinate*. While H$_2$O$_2$ is freely diffusible and continuously replenished, its cellular abundance is low compared with cysteine residues. At 1 μM H$_2$O$_2$ and $\approx 5 \times 10^{10}$ cysteines per cell, there are roughly $10^{-5}$ peroxide molecules per cysteine at any instant. The probability of oxidation depends on the transient coincidence of a peroxide trajectory with the local field envelope surrounding a reactive cysteine. Because the field amplitude decays exponentially with the Debye length ($\lambda\_D \approx 0.8 - 1\ nm$), the limiting step is the encounter with the local EMF.

Even when encounters occur, the magnitude and orientation of the local EMF vary across proteins, conformations, and subcellular environments. Each cysteine samples a unique combination of surrounding charges, dielectric boundaries, and polarization states. Because $k(E)$ depends exponentially on the EMF, small geometric or electrostatic differences translate into orders-of-magnitude variations in apparent rate constants. Two molecules of the same protein can display distinct oxidation kinetics simply by residing in slightly different microenvironments. Oxidation kinetics are not kinetically uniform. Instead, they define a patchwork of rates, producing heterogenous oxidation profiles across the proteome. Redox sensing becomes mutable by design—an evolving network of rate constants tuned by fluctuating electromagnetic landscapes.

By embedding the EMF into biochemical kinetics, the equations of life become continuous with Maxwell's and Planck's laws. Rate constants cease to be immutable numbers—they are field-conditioned expressions of geometry, charge, and time[15]. This realisation reframes biochemical reactivity as an emergent property of the local EMF. Slow chemistry in bulk solution becomes fast enough to sustain signalling, metabolism, and adaptation.

**Methods**

**1. Pseudo–first-order bulk kinetic calculations.** Based on the literature[3,7], cysteine oxidation by $H_2O_2$ was modelled[16] with a representative field-free rate constant $k_2 = 10\ M^{-1}s^{-1}$ pH 7.4, 37°C. Assuming $H_2O_2$ is replenished on a timescale fast relative to reaction progress (constant $[H_2O_2]$) and neglecting reduction during the window[17], the reaction is pseudo–first-order with respect to cysteine:

$$k' = k_2[H_2O_2]$$

For an initially reduced cysteine proteoform[18] population, the time to reach a fractional oxidation $f$ is

$$t_f = -\frac{\ln(1-f)}{k'}$$

The time to reach a target fraction was set to a conservative 5% ($f = 0.05; t_{5\%}$). Under these assumptions, $t_f$ is independent of the protein copy number ($N$), but for completeness this parameter was included. To calculate $t_{5\%}$ in a HeLa cell[19], volume 3 picoliters, the equations were computationally instantiated over $[H_2O_2] = 1\ nM\ to\ 1\mu M$ using a script.

**2. Reverse kinetic calculations.** To identify the conditions necessary to achieve 5%-oxidation in 5-min, the inverse expressions

$$[H_2O_2] = -\frac{\ln(1-f)}{k'}, k_2 = -\frac{\ln(1-f)}{t[H_2O_2]}$$

were solved computationally using a script.

**3. Local electromagnetic field $E_{loc}$.** Local electromagnetic fields were computed by the superimposition of screened sources within the electro-quasistatic (EQS) limit of Maxwell's equations,

$$\nabla \cdot (\varepsilon(\boldsymbol{r})\nabla\phi) - \varepsilon(\boldsymbol{r})\kappa^2\phi = -\rho_f, \boldsymbol{E} = -\nabla\phi$$

where $\varepsilon(\boldsymbol{r})$ is the spatial varying permittivity, $\kappa = 1/\lambda_D$ the inverse Debye length, and $\rho_f$ the charge density of residues and membrane.

Boundary conditions require continuity of tangential E and of normal displacement D:

$$\varepsilon_2 E_{2n} - \varepsilon_1 E_{1n} = \sigma_f$$

For charge-neutral interfaces ($\sigma_f \approx 0$), this yields the dielectric focusing relation

$$E_{in} \approx \frac{\varepsilon_{out}}{\varepsilon_{in}} E_{out}$$

Hence a cysteine in a low-$\varepsilon$ protein pocket ($\varepsilon_{shell}\sim 10$) adjacent to bulk water ($\varepsilon_{bulk}\sim 80$) experiences a normal-field amplification $F_\varepsilon = \varepsilon_{bulk} / \varepsilon_{shell}$.

Screened contributions were modelled as follows:

**Point charge** $q$ at distance $r$

$$E_{pt}(r) = \frac{1}{4\pi\varepsilon_0\varepsilon_r} q \frac{e^{-\kappa r}}{r^2}(1+\kappa r)$$

**Point dipole** $p$, on-axis:

$$E_{dip}(r) \simeq \frac{1}{2\pi\varepsilon_0\varepsilon_r} q \frac{p}{r^3} e^{-\kappa r}(1+\kappa r)$$

**Diffuse layer with** surface potential $\phi_0$

$$E_\perp \approx \frac{\phi_0}{\lambda_D} e^{-z/\lambda_D} \; z \; normal \; distance$$

The total net field along the reaction coordinate is the projected component,

$$\boldsymbol{E}_\parallel = \boldsymbol{E}\cdot\hat{\boldsymbol{n}}$$

**4. Field-aware rate constant.** The local electromagnetic field was embedded into the Eyring equation as

$$k(E) = \kappa_0 \frac{k_B T}{h} exp\left[-\frac{(\Delta G_0^\ddagger - \Delta\mu\, E - \frac{1}{2}\Delta\alpha E^2)}{RT}\right]$$

where $\Delta\mu_\parallel = \Delta\boldsymbol{\mu}\cdot\hat{\boldsymbol{n}}$ is the dipole change projected onto the field, and $\Delta\alpha$ is the polarizability tensor change. The exponential term captures barrier lowering (enthalpic contribution).

**5. Orientational focusing and the reaction cone**. Field alignment confines reactive orientations to a cone of half-angle $\theta_c$ defined by

$$cos\theta_c = L(\xi), \xi = \frac{\mu_{eff} E_{loc}}{k_B T}$$

where $L(\xi)$ is the Langevin function. The probability of being in the cone is

$$P(\theta \leq \theta_c|\xi) = \frac{e^\xi - e^{\xi cos\theta_c}}{e^\xi - e^{-\xi}}, P_0 = \frac{1-cos\theta_c}{2},$$

and the orientational gain

$$G_{orient} = \frac{P}{P_0}$$

The effective prefactor becomes

$$\kappa(E) = \kappa_0 G_{orient} f_{S^-} f_{open}$$

where $f_{S^-}$ is the thiolate fraction and $f_{open}$ any conformational gating.

Thiolate speciation follows from $pK_a^{loc}$

$$f_{S^-} = \frac{1}{1 + 10^{(pK_a^{local} - pH)}}$$

with $pK_a^{loc}$ itself shiftable by Born and field effects in low-$\varepsilon$ microenvironments.

**6. Oscillatory fields.** For $E_t = E_0 \cos \omega t$ with $\omega$ fast relative to reaction times, the time-averaged barrier separates into linear Stark average and a quadratic average

$$\langle e^{(\Delta\mu_\| E)/(RT)} \rangle = I_0 \left( \frac{\Delta\mu_\| E_0}{RT} \right), \langle e^{(\frac{1}{2} E \cdot \Delta\alpha \cdot E)/(RT)} \rangle \approx \exp\left( \frac{E_0^2 \Delta\alpha_\|}{4RT} \right)$$

using $\langle E^2 \rangle = E_0^2/2$ and the modified Bessel function $I_0$.

**7. Summary of the field-aware rate constant.**

$$k(E) = \frac{k_B T}{h} \kappa_0 G_{orient}(\xi, \theta_c) f_{S^-} f_{open} \exp\left[ \frac{\Delta\mu_\| E_\| + \frac{1}{2} E \cdot \Delta\alpha \cdot E - \Delta G_0^\ddagger}{RT} \right]$$

The observed pseudo first-order rate constant is

$$k'(E) = k(E) [reactant]_{local}$$

where $[reactant]_{local}$ is shaped by geometry, dielectric, and nanoconfinement.

**8. Computationally implementing the field-aware rate constant.** To computationally implement the field-aware rate constant, physical units were set to SI; $Debye \rightarrow Columbic \cdot m$ ($3.33564 \times 10^{-30}$) and $Å^3 \rightarrow SI$ polarizability ($4\pi\varepsilon_0 \times 10^{-30} m^3$) with conversions explicitly noted in the [script](script). To quantify local fold acceleration $k(E) / k_0$ from local fields and dielectric focusing, separating barrier lowering and orientational focusing.

$$k(E) = \frac{k_B T}{h} \kappa_0 G_{orient}(\xi, \theta_c) \exp\left[ \frac{\Delta\mu_\| E_\| + \frac{1}{2} E \cdot \Delta\alpha \cdot E - \Delta G_0^\ddagger}{RT} \right],$$

with $\xi = \mu_{eff} E / k_B T$. The script reported the fold acceleration

$$\frac{k(E)}{k_0} = G_{orient}(\xi, \theta_c) \exp\left( \frac{\Delta\mu E + \frac{1}{2} \Delta\alpha E^2}{RT} \right)$$

The script implemented:

- $G_{orient}$. Probability within a reaction cone (half-angle $\theta_c$) under field bias $\propto e^{\xi \cos\theta}$; returned as $P / P_0$ (biased over isotropic).

- Barrier gain $exp[(\Delta\mu\, E + \frac{1}{2}\Delta\alpha E^2) / RT)]$; $\Delta\mu$ in Debye, $\Delta\alpha$ Å³ (optional set to 0 by default; use 40 Å³ to test Stark-polarizability contribution).
- Dielectric focusing. Normal component amplified by $F_\varepsilon = \varepsilon_{bulk} / \varepsilon_{shell}$; code sweeps $F_\varepsilon \in \{1,2,3,4,5\}$ by evaluating $E_{loc} = F_\varepsilon\, E$.
- A parameter sweep $E \in [10^7, 3 \times 10^8]\, V\, m^{-1}$ (log-spaced); $\Delta\mu \in \{10,15,20\}D$; $\mu_{eff} \in \{1,23\}\, D$; $\theta_c \in \{20°, 30°, 45°\}$; $F_\varepsilon \in \{1,2,3,4,5\}$. Baseline $k_0 = 10\, M^{-1}s^{-1}$.

The parameter sweep data are freely available online.

To translate fold enhanced $k_2$ into observable timescales for a pseudo-first-order regime at fixed $[H_2O_2]$, a script was implemented. With kinetics as $k' = k_2[H_2O_2]$, the time to fraction $f$ remained

$$t_f = -\frac{\ln(1-f)}{k'}$$

The script computed $f = 0.05$ (5%). The code implemented a parameter sweep: $[H_2O_2] \in \{0.001 - 1\mu M\}$ (log-spaced); baseline $k_2 = 10\, M^{-1}s^{-1}$; fold-enhancements $\{1,3,10,30,100,300,1000\}$. The parameter sweep data of doses and enhancements are freely available online.

**9. Estimating the local EMF near a charged membrane**. To estimate the magnitude and spatial decay of the $E_{loc}$ experienced by a cysteine residue situated near a negatively charged plasma membrane and surrounded by nearby positive amino acid sidechains, under physiological ionic strength, a script was implemented.

The total field at the reactive coordinate was computed as the superimposition of (i) a screened diffuse-layer field arising from a charged membrane surface potential $\phi_0$, and (ii) screened Coulomb fields from neighbouring residue charges. Both were evaluated under Debye–Hückel screening and corrected for dielectric focusing in a low-permittivity protein microenvironment ($\varepsilon_{shell} \ll \varepsilon_{bulk}$).

The script computed the diffuse layer field normal component

$$E_\perp \approx \frac{\phi_0}{\lambda_D} e^{-z/\lambda_D}$$

where $z$ is the distance from the membrane, $\lambda_D$ the Debye length, and

$$\lambda_D = \sqrt{\frac{\varepsilon_r \varepsilon_0 k_B T}{2N_a e^2 I \times 1000}},$$

with $I = 0.15\, M, T = 310\, K$.

Each nearby positively charged arginine or lysine residue was represented as a screened charge per section 3. Vector components were summed and projected along the coordinate as normal.

The code computed the local EMF using the following parameters: $\phi_0 = -70\ mV$; $\varepsilon_{bulk} = 80$; $\varepsilon_{bulk} = 10$; $F_\varepsilon \approx 8$ with three positively charged residues positioned at (0.6,0.0,0.2), (−0.3,0.5,−0.1)(-0.3,0.5,-0.1) nm relative to cysteine. The code reproduces the field magnitude and decay shown in the main text.

**10. Code availability**. The python-scripted source code is available at https://github.com/JamesCobley/EMF_Redox/tree/main, inclusive of a readme and requirements files. The self-contained code is ready to run on a standard google colab runtime. The codebase is freely available under an MIT license, permitting reuse and modification as desired, provided it is cited via the Zendo DOI.